# An Internet-enabled technology to support Evolutionary Design[1]

V.V. Kryssanov*, H. Tamaki, and K. Ueda

**Abstract**

This paper discusses the systematic use of product feedback information to support life-cycle design approaches and provides guidelines for developing a design at both the product and the system levels. Design activities are surveyed in the light of the product life cycle, and the design information flow is interpreted from a semiotic perspective. The natural evolution of a design is considered, the notion of design expectations is introduced, and the importance of evaluation of these expectations in dynamic environments is argued. Possible strategies for reconciliation of the expectations and environmental factors are described. An Internet-enabled technology is proposed to monitor product functionality, usage, and operational environment and supply the designer with relevant information. A pilot study of assessing design expectations of a refrigerator is outlined, and conclusions are drawn.

<u>Keywords</u>: Evolutionary design, product life cycle, Internet.

* The corresponding author. Inquiries should be send to the following address: Dr. Kryssanov V.V., DEPT. Computer and Systems Engineering, Faculty of Engineering, Kobe University, Rokko-dai, Nada-ku, Kobe 657-8501, JAPAN. E-mail: kryssanov@al.cs.kobe-u.ac.jp

---

# 1 Introduction

Modern society places severe demands on the industry to increase its efficiency but reduce the environmental impact caused by production activities. Besides, there is a widening awareness that to protect the environment and save the natural resources for future generations, it is not enough to merely advance and optimize the processes in design, planning, machining, etc., which are directly associated with the development of products [10]. Companies striving for leadership in the modern market place more and more realize the necessity of enlightening customers on recycling and other relevant life-cycle issues, influencing legislation, and proactively shaping in this way the future market. To sustain these industrial needs and bring environmental considerations into production processes, the focus of design paradigms has recently moved from end-of-pipe solutions to complex life-cycle approaches [1], [20].

In the past few decades, a substantial amount of knowledge on how to minimize the environmental impact throughout the product life cycle has been accumulated in academia and the industry, and several 'environmentally conscious' design strategies have been proposed. Cleaner Production, Green Design, Design for Environment, Eco-design – this is by far not an exhaustive list of so-called life-cycle approaches developed by the design community. Struggling with integration of different strategies of all life cycle stages into one, these approaches provide with powerful methods for introducing environmental aspects in product development. However, every of the life-cycle strategies proposed proceeds from having a product already fabricated, and it is based on an extensive analysis of product information, which requires considerable expertise and time to collect and process, as well as to interpret [10]. Another drawback is that the popular life-cycle design strategies stimulate incremental changes of eco-efficiency at the product level only, i.e. in terms of the product, its structure, constituents, and functioning, – they cannot be applied at a more general system level, which is broadly defined on product environments (technical, technological, social, etc.) and in terms of product-environment interactions [18]. The latter makes these strategies fundamentally limited

The aim of the presented study is to develop a methodology of evolutionary design support that would allow one to overcome certain of the major difficulties in practical implementation of the life-cycle design approaches. The methodology is based on a new semiotic paradigm of evolutionary design, and it allows one to empirically evaluate design requirements in post-production life cycle stages and provide guidelines for evolving a design at both the product and the system levels. The main premise of the approach is that any design fits to certain, once fixed requirements of continuously developing reality. Reality cannot, however, be fixed, and the realized design requirements can fail if pushed too far. Hence, the natural evolution of a design as an information object should reflect the change of the requirements.

The study begins from reviewing design activities from an information processing perspective. The life cycle of a product is described, and the natural evolution of a design as an information object throughout the life cycle is considered. A semiotic model of the design evolution is proposed. In Section 3, the notion of design expectations is introduced. The importance of reconciliation of design expectations and the corresponding parameters of product environments is argued. A classification of design expectations is proposed, and the problem of evaluation of design requirements in dynamic environments is discussed. Section 4 presents the concept of expectation agents – software-hardware units, in which design expectations are encoded in the form of programmable agents that can monitor product functionality, usage, and operation environment and perform various control actions when the expectations are violated. A case study of the deployment of an expectation agent developed to monitor a refrigerator is described in Section 5. The agent is applied to collect design expectations -related information using the existing information infrastructure of the Internet. The study conclusions are finally given in Section 6.

## 2 Product life cycle and the design information flow

In the literature, the life cycle of a product is typically represented in the form of a circle displaying the (closed) loop of the material flow (see [2], [20]). Such a representation makes focus on physical processes requisite for the development, use, and retirement of a product, and the task of a life cycle approach to design is usually formulated as to increase the efficiency of the processes but minimize their (negative) influence on the environment. While this view seems quite appropriate for optimization in the post-design stages, it is hardly applicable to the design process itself. Indeed, any design is a description of a product that not yet exists, and the task of design is the task of the creation of such a description. Therefore, design is, first of all, an information processing activity that cannot fully be described in terms of physical entities. The purpose of this section is to define and interpret the flow of design information in a context of the product life cycle.

**2.1 The natural evolution of a design of a product**

Figure 1 gives a simplified view of the design information flow in the life cycle of a product and shows the evolution of the product. The cycle begins from need recognition and extends through planning, design, production, logistics, utilization, maintenance, and finally, product removal, which can be preceded by recovery and re-manufacturing and includes recycling and disposal. The life cycle starts from gathering marketing information, its analysis, and elaboration of requirements necessary for the development of a (new) product concept. Once the requirements have been conceived, they are translated into product functions, a suitable product structure is found (adopted from past experience or newly generated), and product components are determined and related to the functions. A conceptual layout of the product is thus created and then evaluated in accordance with technical, economic, ecological, ethic, etc. criteria. Product specifications are refined, impracticable (under given conditions) design alternatives are eliminated, the design space is fixed, and design local optima with best performances and least risks are sought. Completing the design stage, a single global solution is chosen, detailed, and optimized. The resultant design is transformed into its process plan with a schedule, and the product can be prototyped to check its manufacturability and validate the technological solutions. The entire product concept is validated at the post-production stages. Customers and support service technicians' experience of dealing with the product as well as empirical data of its marketability, operation in an environment, and disposal effects direct the design evolution in next versions of the product.

Design is a progressive, purposeful, and finite process in the sense that it is sequenced in time from the initial concept to the completed product. Besides, design is nonlinear and continual: the cognitive processes responsible for designing are hardly ordered, and design thought freely moves from one aspect of the problem to another (for instance, driven by feedback from the customers) [17]. Design can also be multileveled (with several parallel activities) and discontinuous (in both time and space). This process is not always successful in the sense of yielding the product, while its objectives may be abandoned or redefined arbitrarily at almost any stage of the product life cycle [10].

The design process is frequently thought of in terms of several sequential phases: conceptual design, preliminary or embodiment design (layout), and detailed design. Transitions from one phase to another have a 'quantum', emergent, and bi-directional character and depend on available information and knowledge. The internal development cycle within each phase of the design process generally includes problem formulation, generation of solutions, evaluation, and decision. A transition between phases is made when no external information (in respect to the current design space) is needed for decision at the given phase or when available information is not enough for solution generation. The process of the whole design evolution has a similar structure and dynamics – a decision based on life-cycle information may initiate a quantum transition to consideration of another version of the product (see Figure 1 that depicts the major flows of design information).

## 2.2 Design as a semiosis process

Due to its non-monotonicity and unpredictability at the system level, the process of design (and, therefore, product) evolution can hardly be described by a fixed single model. This process can, however, be described by a metamodel, representing the (possible) transitions of one model to another model, as the general theory of complex system evolution prescribes [8].

In [13], [14], the authors have formulated the basics of a new semiotic theory of evolutionary design that provides a powerful formal tool for studying the natural evolution of a design as an information object. The theory investigates the interaction of three abstract subjects – the *sign*, its *object*, and its *interpreter* – in a context of the product life cycle and treats design as a semiosis process [7]. The theory argues that no truthful description of reality can be made outside the limits of human perception, and during designing, each designer constructs a unique representation of the problem that, nonetheless, is subject to the common laws of design semiosis.

By the semiotic approach, human cognition at any stage of the product life cycle is characterized as a structuring of experience and perception to provide structured information – a (not necessarily verbal) language – that is to deal with the product. A design is seen as a text 'written' in such a product language with a *syntax*, which constrains the product's topological organization, *semantics*, which mainly defines the product-environment interaction, and *pragmatics*, which manifests physiological, psychological, and sociological effects associated with the product. This language is composed of signs and is, in itself, a *sign system* that influences and, conditional on the usage context, determines the meaning conveyed with signs. The sense made of a sign (i.e. the *interpretant*) is understood as also a sign in the interpreter's mind.

At every stage of its development, a product is perceived through (the manifestation of) its distinctions, which are revealed as technologic, contextual, and ergonomic relations between product parts and between the product and its environment (that includes consumers). The relations are represented in a language. Every time, the language (that is a sign system) may be different but has a common ground – the reality – that fundamentally constrains the relations, provided that human perception is (relatively) uniform and consistent. The language does not prescribe a model, but instead forms a universe of models for physical and mental phenomena of reality that allows designers to reach different solutions while dealing with the same problem. The laws of design semiosis govern the evolution of the product language, which defines the product through its distinctions at both ontological (mereological, topological, morphological, teleological) and epistemological (psychological, aesthetic, semiological) levels of description, throughout the product life cycle [18].

Consumers' needs are originally detected as stable patterns of conceptual relations. An initial product concept can be seen as a composition of related signs. The evolution of the concept in the product life cycle can be described in terms of transitions between different relation patterns by means of a distinction dynamics [8]. This dynamics is based on the interaction of the processes of variation and selection that results in an invariant distinction and constrains the variety of the current product concept. For a design of a product intended to operate in an environment, its variety is determined by the devised product structure (i.e. the relations established between product parts) and the possible relations between the product and the environment (i.e. the product feasible states), which together aggregate the product possible configurations. The variety is defined on and in terms of the product language that includes elements for description of both the structure and environment. Generally, no distinction change can be predicted in advance, and to cope with the changing of the variety, the distinction dynamics forces the evolution of product language that thus reflects this dynamics. The evolution is driven by variation that goes through different configurations of the product and eventually discovers (by selection at every stage of the product life cycle) configurations, which are stable. A constraint on the configurations is then imposed, resulting in the selective retention that decreases the variety but specializes the product language so that only conceptual relations fitting to the environment (i.e. stable relation patterns) ultimately remain. Hence, a fundamental principle – a law – of design semiosis is that the distinctions dynamics normally seeks to decrease the number of possible relations between the product and its environment; in other words, *the most probable direction of the product (concept) evolution is from complex to simple interface*. Technological, economical, and social considerations dictate the need for the development of products, which could maintain in many and various environmental situations. This allows us to complement the principle of product evolution by an assertion that *product structure is naturally evolved from simple to complex* (also, see Section 3.2).

It is interesting to note that this model of design evolution explicates a well-known observation that a really new product usually has a simple structure, but it is difficult in operation because of many possible relations between the product and its environment. As the product matures, its structure becomes complex, but its operation is simplified under the natural evolution. Another instance that well illustrates the fundamental law of design semiosis is the recent introduction of the concept of 'features' (that are, in the case of form features, stable relation patterns defined on the geometrical primitives) into CAD/CAPP practice (see [12]).

It is important to understand that, although the above pattern is valid for both the product and the system levels of design evolution, it only shows the most probable direction of the evolution – from simple to complex structure, but it does not ban a backward dynamics of the variety, i.e. structure simplification. The direction in design evolution depends solely on characteristics of the product environment. A new version of the product emerges when incremental changes in product design change the originally conceived product configuration or when the environment changes, enforcing changing the product language. While life-cycle design approaches traditionally focus on the first factor – the changing of the configuration, the problem of evaluation of the product environment and adaptation of the product design (and, therefore, the corresponding product language/sign system) to the environment is still open. An approach addressing the latter problem is described in the following sections.

## 3 Design expectations

### 3.1 Expectations in the design process

In reality, the number of product behavior scenarios, which can take place throughout the product life cycle, is infinite. However, obviously, a design as a model can sustain only a finite set of such scenarios. During designing, designers have to consider various expectations about how the product would behave and interact with the environment. The expectations are revealed as stable relation patterns defined in the product language (that can, at the applied level, be understood as a design grammar) and, as we shall further see, they critically affect all the post-design stages of the product life cycle. Sources of knowledge for these expectations (called design expectations) can be specifications of similar products, knowledge of the designed product domain, knowledge of the specific tasks and working environments of product users, relevant past experience, and so on. In many cases, service and maintenance companies collaborate with manufacturers as sub-contractors and provide a product usage history and failure statistics to the designers. Some design expectations can explicitly be represented (for instance, as pre-defined instructions on the typical usage/service scenarios), while others are assumed 'by default' and remain implicit until they are realized at a later stage of the product life cycle (e.g. the assumption that the product will be used by only a right-handed person).

A product cannot be isolated from its settings, and therefore, a design model of the product must describe not only its functionality, but to a considerable extent the assumed product-user and product-environment interactions. In [5], the authors defined three main groups of design expectations. Depending on the origin of the corresponding relation pattern, design expectations can be classified as: 1) *functional* – the product is expected to properly function if during its operation, no failure occurs and all of the functional parameters are kept within the designed range of tolerances, 2) *environmental* – environmental conditions are meant to be predictable if all the design parameters describing the product operational environment are within the conceived range of tolerances; detecting new parameters, i.e. emergence of new relation patterns that strongly affect the product behavior and/or characteristics is also considered a violation of the environmental expectations; and 3) *usage-related* that include expectations about the user-product interactions.

If the environment (that includes customers), in which a product operates, behaves in an unexpected way, the product may fail or exhibit functional characteristics outside the pre-established range of product tolerances. Detecting and resolving mismatches between the designer's expectations and the actual practice is important for improving product usability, preventing from product premature withdrawal and the connected economic loses, and protecting the environment. When detected, the mismatches can be corrected in two general ways. To better address the environmental factors – customer needs, legislation, market dynamics, etc. – designers can modify their expectations (and the corresponding requirements) by appropriately correcting that, which we call the product language (or, synonymously, sign system), i.e. (the representations of) basic concepts and principles for creating the design, and re-designing the product. Alternatively, contradictory elements of the given environment can be adapted or replaced to secure the product operation (for instance, customers can be encouraged to learn more about the developer's expectations and to correctly operate the existing product).

From the standpoint of design semiosis, all the design expectations are defined in terms of a sign system, while any violation of the expectations activates the evolution of this system.

Consequently, realizing evolutionary design at the level of computer applications requires solving two major problems: i) to computationally define the sign system (i.e. product language), and ii) to ensure (to analyze) its evolution, based on product feedback information. The former is essentially of the development of design grammars – a traditional and quite well established research area in the fields of CAD and AI (see [3]). The latter – utilization of product feedback information – remains, however, poorly explored and receive little computer support in engineering design. Through the presented study, the authors have developed an agent-based Internet-enabled technology, which is described in Section 4, that realizes the main laws of design semiosis and allows for supporting evolutionary design of high-tech products with computers (also, see [6]).

### 3.2 Dynamic environments

To ensure the customer's satisfaction and improve eco-efficiency of the life-cycle processes, the designer has to utilize feedback information from all the product life cycle stages. This is usually done through analyzing empirical data of handling products similar to the one under consideration and/or through evaluating the experience of yielding a product development batch. Various statistical methods and monitoring techniques can then be used [4]. The purpose of the designer is formulated as follows: based on available knowledge and product feedback information, find a product configuration that is intrinsically stable and can adequately react to changes in the intended environment.

It should be noted however, that any feedback information is necessarily local and subject to change, since product environments (social, technical, operational, etc.) continually evolve. Every design is based on expectations that fit particular conditions, which are likely to become obsolete before the product reaches into the market place. From the standpoint of natural evolution of complex systems [8], a universal approach to solving this problem would be to increase as much as possible the internal variety of the product by contriving appropriate decisions in design. Indeed, the more elaborate structure of the product, the larger the number of environmental situations in which it can maintain. Different product configurations would fit (or be adapted to) different situations and thereof, in the case of dynamic environments, design evolution should increase the internal variety, to make the product more complex. Whereas the latter statement is true in general, this does not mean that the 'best' product must be the most complex one. Rather, due many reasons – economical (costs), technical (reliability), ecological (energy/material consumption, pollution), social and ergonomic (safety, convenience and easiness in production and operation), etc., the best would be a product with the simplest possible structure for the given functionality, i.e. with the least possible (for the given environment) internal variety. In traditional design paradigms, this dilemma of balancing the variety is approached as a task of design multi-objective optimization that is always hard, if at all possible, to accomplish in actual manufacturing and that has little to do with the design process as a human activity [17]. Another way would be to not only allow for changing design expectations at the abstract level of product language during designing, but to devise methods of adjusting design expectations at the physical level (e.g. by changing the product configuration, operating mode, maintenance routine, etc.) after the product have been fabricated. For the modern high tech products, the latter can be realized with so-called expectation agents.

## 4 Expectation agents

Elaborating the idea of expectation-driven event monitoring that was originally developed for the software engineering needs [9], an agent-based approach to product usage monitoring has been proposed to gather and utilize design feedback information [5], [15]. It was suggested to encode design expectations in the form of programmable agents, called expectation agents, which monitor product functionality, usage, and operational environment. Data obtained with the agents can be used to optimize the existing product configuration in design, and to detect when usage and environmental patterns shift, thereby necessitating modifying the current product language and re-designing the product. The agents are also to execute various control actions and perform preliminary data processing as well as to provide guidance or suggestions to users and even collect feedback directly from users. Naturally, expectation agents could be used to adjust some of the design expectations materialized in the monitored product.

In the agent-based approach, monitoring units that are presently used for the purposes of condition-based maintenance can be employed (see [16]). Such units typically consist of hardware and software parts. Hardware includes transducers (sensors), pre-processing blocks, interfaces, and programmable blocks. Often, the unit program code is implemented on chips, which cannot be re-programmed, but advantages of re-programmable software-based units are evident. Software-based expectation agents are easy to update/upgrade, and they can be used to supervise the product operation, execute control actions, and generate new events in the product environment.

The agent registers events concerned with an unexpected functional behavior (malfunction), and unexpected environmental conditions, including the user's unexpected actions. An important function of the agent is reporting the event data back to the designers. Among other functions, alarming and notifying the user and the designer about mismatches, data and event history logging, and action control should be pointed to in the first place. Logically, *responding to violations of functional expectations should propel design evolution at the product level*, while *violations of environmental expectations should activate design evolution at the system level*. To adjust design expectations, the agents would change the internal variety of the product by changing the product structure (e.g. at the logical level, if the product consists of electronic blocks) and/or regulate the number of possible relations between the product and other components in the environment (e.g. by modifying the product interface or operation mode).

There are a few important issues in the implementation of expectation agents. To avoid overload of information channels and improve the quality of feedback, data filtration is required such that only filtered, aggregated, and compressed events are transmitted. Instead of reporting every event that has occurred, software-based agents should recognize principal patterns of events affecting the product (e.g. malfunctions) and then derive higher-level control actions and events. Hence, expectation agents must have a function of performing event abstraction to pre-process large volumes of raw sensory information to ensure the transferal of much less data to the higher (decision-making or control) level. For the purposes of analysis, an unexpected situation detected by a monitoring agent should be logged as a record containing at least an identifier of the situation, values of the monitoring parameters, and a time stamp of the occurrence. (Naturally, expectation agents must be sufficiently autonomous to operate after any failure of the monitored product.) Another serious issue is the security of data circulation between the product and the production company. Because of the problem of data property, data crypto- encoding and decoding should be additional functions of the agents. The World Wide Web structure can be used to transfer design information and connect geographically dispersed scenes of action in the product life cycle.

It should be noted that, due to the complex nature of a modern high tech product and the huge number of its characteristics, it would hardly be possible to encode and encapsulate all design expectations into the agents. This can, nevertheless, be subdued by using simple prognostic techniques that require a limited number of indirect parameters to extrapolate and predict the rest of the characteristics needed to evaluate (see [4] for a relevant forecasting method).

**5 A pilot study**

To explore the main theoretical assumptions and check the applicability of the proposed approach, a simple design grammar (i.e. a product language) reflecting the component and function structure of a refrigerator has been developed, and several expectation agents have been implemented and deployed in an experimental modular setup (Figure 2). The setup has been implemented on the basis of the Virtual Instrumentation architecture (National Instruments, Inc.). Its hardware consists of a PC, a 16-channel modular conditioning system, a 16-channel digital acquisition board, IEEE 488 and RS-232C interfaces for data acquisition, and a LAN board for transferring data to remote users. The installed set of transducers can monitor typical parameters of consumer appliances, including the electric current signature, internal and ambient temperatures, vibration level, and sound intensity. An event-driven agent has been implemented as a Virtual Instrument (a G-application running under the LabVIEW environment). To send the monitoring data to a remote client (e.g. the designer's workstation) by way of the Internet, the G-application is activated as a server-side CGI-application of an HTTP server working in the LabVIEW environment. Client functions of the standard FTP and e-mail protocols have been added to the Virtual Instrument. Whenever needed, the application can be replaced by other G-applications supplied from the designer's site to the remote HTTP and FTP server.

It was assumed that some parts and circuits of the monitored refrigerator would be 'weak' from the standpoint of its usage, maintenance, or reliability. It was also known *a priori* that the most common reasons of repairs for the used refrigerator type are compressor problems and refrigerant leakages.

Design expectations have been represented in the form of 'condition-action' rules. Each of these rules compares the current value of a monitored parameter with a threshold. As many important parameters of the refrigerator are not made known by the manufacturer, the threshold values have experimentally been determined through the study. The event-driven monitoring agent has been developed to deal with such parameters as the ratio of working and idle refrigerator cycles, the maximal permissible outdoor humidity, the maximal permissible duration of door opening, and some others.

Let us introduce the following notation:

$h$ – the outdoor humidity,

$h_{max}$ – the maximal permissible outdoor humidity,

$t$ – current time,

$t_{door\_open}$ – the duration of door opening,

$t_{max}$ – the maximal permissible duration of door opening,

$T_{idle}$ – the idle refrigerator cycle duration,

$T_{work}$ – the working refrigerator cycle duration,

U_EB – an identifier of an unexpected environmental behavior,

U_FB – an identifier of an unexpected functional behavior,

U_UB – an identifier of an unexpected user behavior.

Let us define LOG (U_xx, t) the simplest action of the agent as a procedure recording the current time and the identifier U_xx of the unexpected functional, environmental or user behavior. Some of the elementary 'condition-action' rules of the agent can then be represented as follows:

If ($t_{door\_open} > t_{max}$) then LOG (UUB, t);

If ($T_{work}/T_{idle} > X$) then LOG (UFB, t);

If ($h > h_{max}$) then LOG (UEB, t),

where X is a threshold value.

Analogously, if a monitored parameter P (for example, the power) exceeds its critical value $P_{max}$ (that may indicate a serious system failure), some procedures for urgent actions – let us call them as ALARM actions – can be defined:

If ($P > P_{max}$) then ALARM (U_xx, t).

Preventive maintenance based on alarm notifications can reduce the probability of failures, avert the negative environmental impact, and minimize costs associated with product repairs.

Some control actions have been simulated with the software only, as the monitored product does not allow for arranging external control. (It is understood that a specialized programmable device integrated with the refrigerator could easily replace the product-side PC used in the pilot study.)

The main component of the system – the product and the agent's hardware/software – has been installed at the Maintenance Engineering Laboratory, the University of Tokyo in 1999, while the remote HTTP/FTP-based client software have been allocated at the Faculty of Engineering, Kobe University. The front panel of the Virtual Instrument could remotely be observed at the client sites by the use of a standard Web browse (see Figure 3).

Through an analysis of the feedback information, it has been found that the electric power consumption is the main parameter describing the usage history of the refrigerator. Besides, it has been found that leakage from the refrigerator condenser or evaporator can be estimated through measurements of indirect parameters – the electric current signature and temperature. Processing data of the internal and ambient temperatures, the agent could simulate optimization of the power consumption by changing the ratio of working and idle refrigerator cycles. Another statistics of interest is the history of opening door for each of the refrigerator chambers that provides information on the optimal configuration and/or operating mode of the refrigerator for a given environment. (For instance, it appears natural from the standpoint of design semiosis that the concept of multi-chamber refrigerator could emerge as a result of detection and the structural realization – 'interface simplification' – of the stable relation pattern 'door opening-freezer compartment opening'.)

## 6 Summary

The main contribution of this paper is the expectation-driven agent-based approach to assessing design requirements in post-design life cycle stages that allows for overcoming certain of the principal drawbacks of lifecycle design strategies. The proposed approach has a solid theoretical foundation – the theory of complex system evolution and the semiotic theory of evolutionary design. It integrates the processes of design optimization and innovation into one framework and is potentially applicable for analyzing a mix of several products to estimate the so-called rebound effects of design solutions. A pilot study has been made, and its results have shown us that no significant investments are needed to implement and deploy an expectation agent, and that much of commercially available software could be used to support the agent-based approach. Further research is, however, required to develop appropriate methods and tools to properly utilize multifarious feedback information collected. Presently, the authors plan a large-scale experiment, where an artificial neural network will be used at the client's site to recognize new emergent relation patterns in the product language and promote the design evolution.

**Acknowledgement**

This research has been made within the 'Methodology of Emergent Synthesis' research project (No 96P00702) in the Program 'Research for the future' of the Japan Society for the Promotion of Science. The authors are very grateful to Dr. I. Goncharenko for his significant contribution of the research and development.

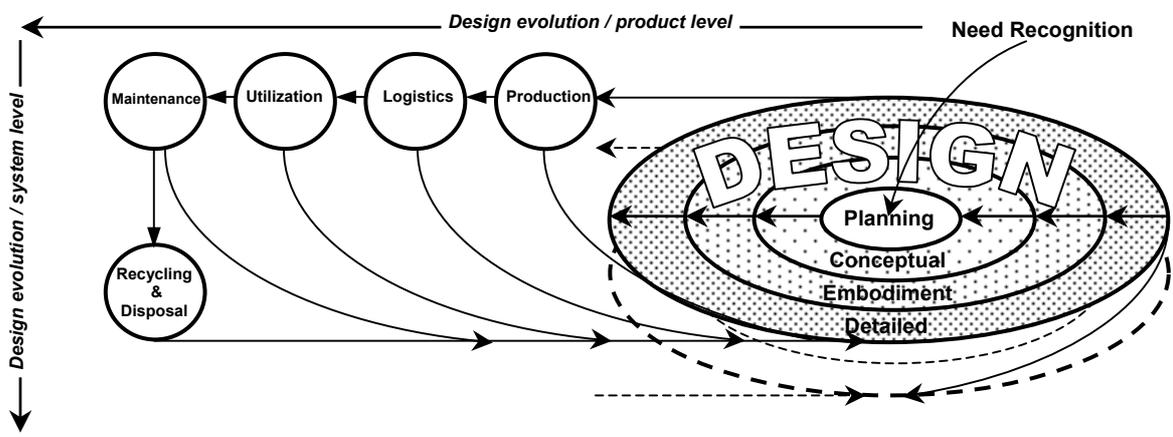

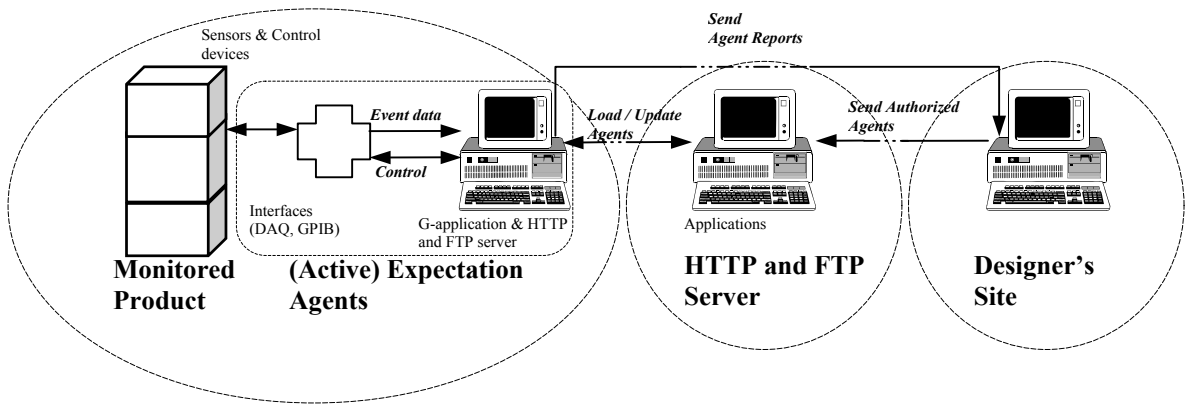

## Monitoring

Vibration: 0.61 Max Vrms2   50.78 Max Freq

Temperature: 24.79

Humidity: 41.56

Noise level, dB: 62.63

### Electric Current

Power, W: 89.0

Wh: 7.664

A Range (AUTO:0): 2

Integration: START

Integrator Time: 0   6   27

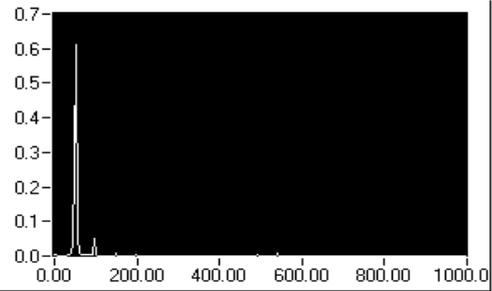

Vibration Spectrum [Vrms2]

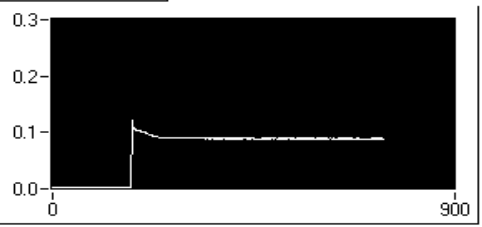

Current Power, kWt

Stop Measurements: STOP

Number of measurements: 742

Figure captions:

Figure 1. Design information flow.

Figure 2. System architecture.

Figure 3. Panel of the Virtual Instrument.